\documentclass[twocolumn,showpacs,preprintnumbers,superscriptaddress,amsmath,amssymb]{revtex4}

\usepackage{graphicx}
\usepackage{dcolumn}
\usepackage{bm}

\begin{document}

\title{From regular to growing small-world networks} 

\author{Zhongzhi Zhang}
\email{zhangzz@fudan.edu.cn}

\author{Shuigeng Zhou}
\email{sgzhou@fudan.edu.cn}
\homepage{http://www.iipl.fudan.edu.cn/~zhousg/sgzhou.htm}

\author{Zhen Shen}

\affiliation{Department of Computer Science and Engineering, Fudan
University, Shanghai 200433, China}%

\affiliation{Shanghai Key Lab of Intelligent Information Processing,
Fudan University, Shanghai 200433, China}

\date{\today}

\begin{abstract}
We propose a growing model which interpolates between
one-dimensional regular lattice and small-world networks. The model
undergoes an interesting phase transition from large to small world.
We investigate the structural properties by both theoretical
predictions and numerical simulations. Our growing model is a
complementarity for the famous static WS network model.
\end{abstract}



\maketitle


\section{Introduction}

Many real-life systems display both a high degree of local
clustering and the small-world
effect~\cite{Ne00,AlBa02,DoMe02,Ne03,BoLaMoChHw06}. Local clustering
characterizes the tendency of groups of nodes to be all connected to
each other, while the small-world effect describes the property that
any two nodes in the system can be connected by relatively short
paths. Networks with these two characteristics are called
small-world networks.

In the past few years, a number of models have been proposed to
describe real-life small-world networks. The first and the most
widely-studied model is the simple and attractive small-world
network model of Watts and Strogatz (WS model)~\cite{WaSt98}, which
triggered a sharp interest in the studies of the different
properties of small-world networks and WS
model~\cite{BaAm99,NeWa99a,BaWi00,AmScBaSt00,Kl00,CoOzPe00,CoSa02}.
The WS model is probably a reasonable illustration of how a
small-world network is shaped. However, the small-world effect is
much more general, researchers begin to explore other mechanisms
producing small-world networks.  Recently, Ozik, Hunt and Ott have
introduced a simple evolution model (OHO model) of growing
small-world networks with geographical attachment preference, where
new nodes are linked to geographically nearby ones~\cite{OzHuOt04}.
A deterministic version of a special case of the OHO model was
presented in~\cite{ZhRoGo05}, and further expanded
in~\cite{ZhRoCo05a}. In addition, many authors found that
small-world properties can be also created in other interesting
ways~\cite{BlKr05,YaHo05}.

 It is well-known that the WS model is the
first successful model that interpolates between a regular ring
lattice and a completely random network, and plays an important role
in network science. However, the WS model is static (i.e. the
network size is fixed), this does not agree with the growth property
of many real-life systems~\cite{BaAl99}. In addition, in real
systems, a series of microscopic events shape the network evolution,
including addition or removal of a node and addition or removal of
an edge~\cite{AlBa00,DoMe00b,ChSh04,WaWaHuYaQu05,ShLiZhZh06}.
Therefore, it is interesting to establish a growing small-world
model to investigate the effect of local events on the topological
features like the WS model. To our best knowledge, all previous
models of small-world networks either are static or only took into
account the addition of nodes and edges without considering other
microscopic events.

In this paper, we present a growing small-world network model
controlled by a tunable parameter $q$, where existing edges can be
removed. By tuning parameter $q$, the model undergoes a phase
transition from large to small worlds as the WS model. We study
analytically and numerically the structural characteristics, all of
which depend on parameter $q$


\section{The model}
\begin{figure}
\begin{center}
\includegraphics[width=0.35\textwidth]{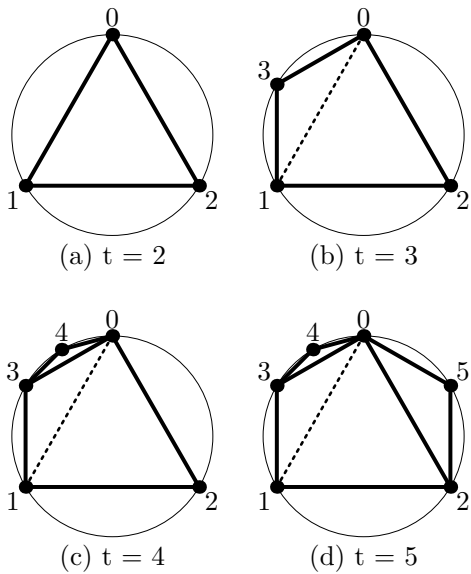}
\end{center}
\caption[kurzform]{\label{fig1} Scheme of the growing network for
the case of $q=0.3$, showing the first four time steps of the
evolving process. The dashed line represents the edge removed.}
\end{figure}

In this section, we introduce a growing model which describes
networks from regular to small-world. The model is constructed in
the following way (see Fig. \ref{fig1}).

(i) \emph{Initial condition}: We start from an initial state ($t=2$)
of three nodes distributed on a ring, all of which form a triangle.

(ii) \emph{Growth}:  At each increment of time, a new node is added
which is placed in a randomly chosen internode interval along the
ring. Then we perform the following two operations.

(iii) \emph{Addition of edges}: The new node is connected its two
nearest nodes (one on either side) previously existing. Nearest, in
this case, refers to the number of intervals along the ring.

(iv) \emph{Removal of an edge}: With probability $q$, we remove the
edge linking the two nearest neighbors of the new node.

The growing processes are repeated until the network reaches the
desired size.

When $q=1$, the network is reduced to the one-dimension ring
lattice. For $q=0$, no edge are deleted, the model coincides with a
special case of the OHO model~\cite{OzHuOt04}. Varying $q$ in the
interval (0,1) allows one to study the crossover between the
one-dimension regular lattice and the small-world networks.

By construction, at every step, the number of nodes increases one,
while the average number of edges added is $2-q$. Then we can see
easily at time $t$, the network consists of $t+1$ nodes and average
$(2-q)t+2q-1$ edges. Thus when $t$ is large, the average node degree
at time $t$ is equal approximately to a constant value $4-2q$, which
shows our network is sparse like many real-life networks.

\section{Topological properties}
We focus on the behavior of the topological characteristics, in
terms of the degree distribution, the clustering coefficient, and
the average path length, as a function of the parameter $q$.

\subsection{Degree Distribution}
The degree distribution is one of the most important statistical
characteristics of a network. For $q=1$, all nodes have the same
number of connections 2, the network exhibits a completely
homogeneous degree distribution. Next we focus the case $0\leq q
<1$. In order to conveniently describe the computation of the
network characteristics, we label nodes by their birth times, $s=0,
1, 2,\ldots, t$, and use $p(k, s, t)$ to denote the probability that
at time $t$ a node created at time $s$ has a degree $k$. At time
$t$, there are $t+1$ internode intervals along the ring and each
node has two intervals (one on either side). The master
equation~\cite{DoMeSa00,DoMeSa01} governing the evolution of the
degree distribution of an individual node has the form
\begin{eqnarray}\label{eq1}
p(k,s,t+1)&=&\frac{2(1-q)}{t+1}p(k-1,s,t)\nonumber\\
&+&\left(1-\frac{2(1-q)}{t+1}\right)p(k,s,t)
\end{eqnarray}
with the initial condition, $p(k,s={0,1,2},t=2)=\delta_{k,2}$ and
the boundary one $p(k,t,t)=\delta_{k,2}$. This accounts for two
possibilities for a node: first, with probability
$\frac{2(1-q)}{t+1}$, it may get an extra edge from the new node
while its existing edges remain undeleted, and thus increase its own
degree by 1; and second, with the complimentary probability
$1-\frac{2(1-q)}{t+1}$, the nodes may remain in the former state
with the former degree. It should be noted that Eq. (\ref{eq1}) and
all the following ones are exact for all $t\geq 2$.\\
\indent The total degree distribution of the entire network can be
obtained as
\begin{equation}\label{eq2}
P(k,t)=\frac{1}{t+1}\sum_{s=0}^{t}p(k,s,t)
\end{equation}
Using this and applying $\sum_{s=0}^{t}$ to both sides of Eq.
(\ref{eq1}), we get the following master equation for the degree
distribution:
\begin{eqnarray}\label{eq3}
(t+2)P(k,t+1)-(t+1)P(k,t)\nonumber\\
=2(1-q)P(k-1,t)-2(1-q)P(k,t)+\delta_{k,2}.
\end{eqnarray}
The corresponding stationary equation, i.e., at $t
\longrightarrow\infty$, takes the form
\begin{equation}\label{eq4}
(3-2q)P(k)-(2-2q)P(k-1)=\delta_{k,2}.
\end{equation}
Eq. (\ref{eq4}) implies that $P(k)$ is the solution of the recursive
equation
\begin{equation}\label{eq5}
P(k)=\left\{\begin{array}{lc}
{\displaystyle{\frac{2-2q}{3-2q}P(k-1)} }
& \ \hbox{for}\ k>2\\
{\displaystyle{1/(3-2q)} }
& \ \hbox{for}\  k=2\\
\end{array} \right.
\end{equation}
giving
\begin{eqnarray}\label{eq6}
P(k)=\frac{1}{3-2q}\left(\frac{2-2q}{3-2q}\right)^{k-2}   \space
(k\geq 2),
\end{eqnarray}
which decays exponentially with $k$. For $q=0$, Eq. (\ref{eq6})
recovers the result previously obtained in~\cite{OzHuOt04}. Thus the
resulting network is an exponential network. Note that most
small-world networks including the WS model belong to this
class~\cite{BaWi00,OzHuOt04,ZhRoGo05,ZhRoCo05a}.

\begin{figure}
\begin{center}
\includegraphics[width=0.4\textwidth]{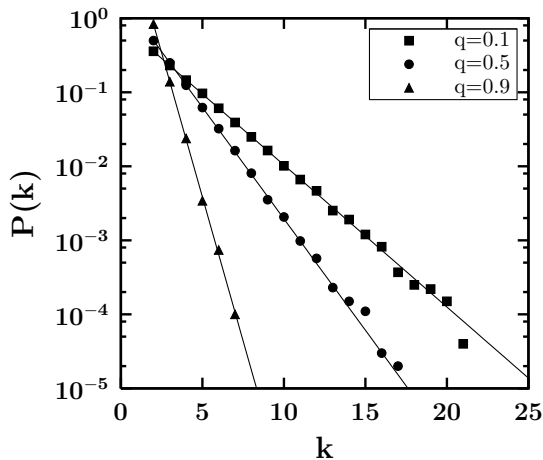}
\end{center}
\caption[kurzform]{\label{fig2} Semilogarithmic graph of degree
distribution of the networks with order $N=10^{5}$. The solid lines
are the analytic calculation values given by Eq. (\ref{eq6}).}
\end{figure}

In Fig.~\ref{fig2}, we report the simulation results of the degree
distribution for several values of $q$.  From Fig.~\ref{fig2}, we
can see that the degree spectrum of the networks is continuous and
the degree distribution decays exponentially for large degree
values, in agreement with the analytical results and supporting a
relatively homogeneous topology similar to most small-world
networks~\cite{BaWi00,OzHuOt04,ZhRoGo05,ZhRoCo05a}.

\subsection{Clustering coefficient}
Most real-life networks show a cluster structure which can be
quantified by the clustering
coefficient~\cite{AlBa02,DoMe02,Ne03,Ne00,BoLaMoChHw06}. The
clustering coefficient of a node gives the relation of connections
of the neighborhood nodes connected to it. By definition, clustering
coefficient $C_{i}$ of a node $i$ is the ratio of the total number
$e_{i}$ of existing edges between all $k_{i}$ its nearest neighbors
and the number $k_{i}(k_{i}-1)/2$ of all possible edges between
them, ie $C_{i}=2e_{i}/[k_{i}(k_{i}-1)]$. The clustering coefficient
$C$ of the whole network is the average of all individual
$C_{i}^{'}s$.\\
\indent For the case of $q=1$, the network is a one-dimensional
chain, the clustering coefficient of an arbitrary node and their
average value are both zero.

For the case of $q=0$, using the connection rules, it is
straightforward to calculate exactly the clustering coefficient of
an arbitrary node and the average value for the network. When a node
$i$ enters the network,  $k_{i}$ and $e_{i}$ are $2$ and $1$,
respectively. After that, if the degree $k_{i}$ increases by one,
then its new neighbor must connect one of its existing neighbors,
i.e. $e_{i}$ increases by one at the same time. Therefore, $e_{i}$
is equal to $k_{i}-1$ for all nodes at all time steps. So there
exists a one-to-one correspondence between the degree of a node and
its clustering. For a node $v$ with degree $k$, the exact expression
for its clustering coefficient is $\frac{2}{k}$, which has been also
been obtained in other
models~\cite{OzHuOt04,ZhRoGo05,ZhRoCo05a,DoGoMe02,HiBe06}. This
expression for the local clustering shows the same inverse
proportionality with the degree as those observed in a variety of
real-life networks~\cite{RaBa03}. In this limiting case, the
clustering coefficient $C$ of the whole network is given by
\begin{eqnarray}\label{eq7}
C=2\sum_{k=2}^{\infty}\frac{1}{k}P(k)=\frac{3}{2}\ln3-1\approx
0.6479.
\end{eqnarray}
So in the limit of large $t$ the clustering coefficient is very
high.

In the range $0<q<1$, it is difficult to derive an analytical
expression for the clustering coefficient either for an arbitrary
node or for the average of them. In order to obtain the result of
the clustering coefficient $C$ of the whole network, we have
performed extensive numerical simulations for the full range of $q$
between 0 and 1. Simulations were performed for system sizes
$10^{5}$, averaging over 20 network samples for each value of $q$.

\begin{figure}
\begin{center}
\includegraphics[width=0.4\textwidth]{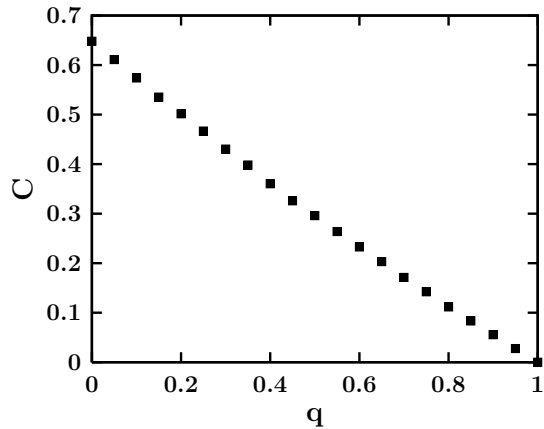}
\end{center}
\caption[kurzform]{\label{fig3} The clustering coefficient $C$ of
the whole network as a function of $q$.}
\end{figure}

In Fig.~\ref{fig3}, we plot the clustering coefficient $C$ as a
function of $q$. It is obvious that $C$ decreases continuously with
increasing $q$. As $q$ increases from 0 to 1, $C$ drops almost
linearly from 0.6479 to 0. Note that although the clustering
coefficient $C$ changes linearly for all $q$, we will show below
that in the large limit of $q$, the average path length changes
exponentially as $q$ . This is little different from the phenomenon
observed in the WS model where $C$ remains practically unchanged in
the process of the network transition to a small world.

\subsection{Average Path Length}
Certainly, the most important property of a small-world network is a
logarithmic average path length (APL) (with the number of nodes).
Here APL means the minimum number of edges connecting a pair of
nodes, averaged over all pairs of nodes. It has obvious implications
for the dynamics of processes taking place on networks. Therefore,
its study has attracted much attention.

For the case of $q=1$, the average path length increases linearly
with network size. For the case of $q=0$, the network grows
stochastically. Generally speaking, for a randomly growing network,
the analytical calculation for APL is difficult. Below, we will give
an upper bound for the APL of this particular case, which shows that
the APL increases at most logarithmically with network size.

If $d(i,j)$ denotes the distance between nodes $i$ and $j$, we
introduce the total distance of the network with size $N$ as
$\sigma(N)$:
\begin{equation}\label{eq8}
\sigma(N) = \sum_{0\leq i< j \leq N-1}d(i,j),
\end{equation}
and we denote the APL by $L(N)$, defined as:
\begin{equation}\label{eq9}
L(N) ={2\sigma(N)\over N(N-1)}.
\end{equation}
In the limiting case $q=0$, the distances between existing node
pairs will not be affected by the addition of new nodes. Then we
have the following equation:
\begin{equation}\label{eq10}
\sigma(N+1) = \sigma(N)+ \sum_{i=0}^{N-1}d(i,N).
\end{equation}
Assume that the node $N$ is added and connected to two nodes
$w_1,w_2$ linked by edge $\mathbb{E}$, then Eq. (\ref{eq10}) can be
rewritten as:
\begin{eqnarray}\label{eq11}
\sigma(N+1) &=& \sigma(N)+ \sum_{i=0}^{N-1}\left[D(i,w)+1\right]\nonumber \\
&=&\sigma(N)+N+ \sum_{i=0}^{N-1}D(i,w),
\end{eqnarray}
where $D(i,w)=\min\{d(i,w_1),d(i,w_2)\}$. Constricting the edge
$\mathbb{E}$ continuously into a single vertex $w$ (here we assume
that $w\equiv w_1$), we have $D(i,w)=d(i,w)$. Since
$d(w_1,w)=d(w_2,w)=0$, Eq. (\ref{eq11}) can be rewritten as:
\begin{equation}\label{eq12}
\sigma(N+1) =\sigma(N)+N+ \sum_{i\in \Gamma}d(i,w),
\end{equation}
where $\Gamma= \{0,1, 2,\cdots ,N-1\}-\{w_1,w_2\}$ is a node set
with cardinality $N-2$. The sum $\sum_{i\in \Gamma}d(i,w)$ can be
considered as the total distance from one node $w$ to all the other
nodes in the network with size $N-1$, which can be roughly evaluated
by mean-field approximation in terms of $L(N-1)$
as~\cite{ZhYaWa05,ZhRoCo05,ZhRoZh06}:
\begin{equation}\label{eq13}
\sum_{i\in \Gamma}d(i,w)\approx (N-2) L(N-1).
\end{equation}
Note that, as $L(N)$ increases monotonously with $N$, it is clear
that:
\begin{equation}\label{eq14}
(N-2)L(N-1) = {2\sigma(N-1) \over N-1} < {2\sigma(N) \over N}.
\end{equation}
Combining Eqs. (\ref{eq12}), (\ref{eq13}) and (\ref{eq14}), one can
obtain the inequation:
\begin{equation}\label{eq15}
\sigma(N+1) < \sigma(N) +N + {2\sigma(N) \over N}.
\end{equation}
Considering Eq. (\ref{eq15}) as an equation and not an inequality,
we can provide an upper for the variation of $\sigma(N)$ as
\begin{equation}\label{eq16}
{d\sigma(N) \over dN} = N + {2\sigma(N) \over N},
\end{equation}
which leads to
\begin{equation}\label{eq17}
\sigma(N) = N^2(\ln N + \alpha),
\end{equation}
where $\alpha$ is a constant. As $\sigma(N) \sim N^2\ln N $, we have
$L(N) \sim \ln N$. Note that as we have deduced Eq. (\ref{eq17})
from an inequality, then $L(N)$ increases at most as $\ln N$ with
$N$. Therefore, we have proved that in the special case of $q=0$,
there is a slow growth of APL with network size $N$.

\begin{figure}
\begin{center}
\includegraphics[width=0.4\textwidth]{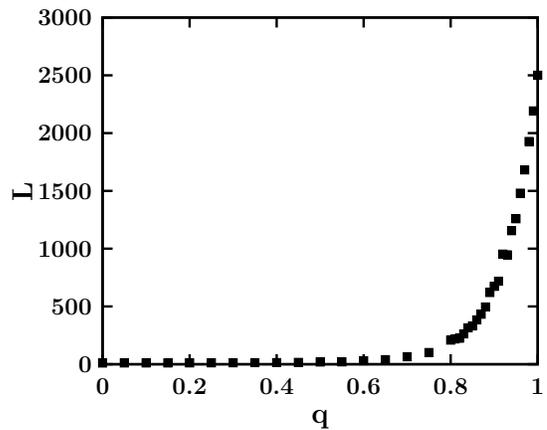}
\end{center}
\caption[kurzform]{\label{fig4} Average path length $L$ versus
parameter $q$.}
\end{figure}

For $0<q<1$, in order to obtain the variation of the average path
length with the parameter $q$, we have performed extensive numerical
simulations for different $q$ between 0 and 1. Simulations were
performed for system sizes $10^{4}$, averaging over 20 network
samples for each value of $q$. In Fig.~\ref{fig4}, we plot the
average path length $L$ as a function of $q$. We observe that, when
lessening $q$ from 1 to 0, average path length $L$ drops drastically
from a very high value a small one, which predicts that a phase
transition from large-world to small-world occurs. This behavior is
similar to that in the WS model.

Why is the average path length $L$ low for small $q$? The
explanation is as follows. The older nodes that had once been
nearest neighbors along the ring are pushed apart as new nodes are
positioned in the interval between them. From Fig.~\ref{fig1} we can
see that when new nodes enter into the networks, the original nodes
are not near but, rather, have many newer nodes inserted between
them. When $q$ is small, the network growth creates enough
"shortcuts" (i.e. long-range edges) attached to old nodes, which
join remote nodes along the ring one another as in the WS
model~\cite{WaSt98}. These shortcuts drastically reduces the average
path length, leading to a small-world behavior.

\section{Conclusions}
In summary, we have proposed a one-parameter model of growing
small-world networks. In our model, in addition to new edges
connecting new nodes and old ones, edges between old nodes may be
removed. The presented model interpolates between one-dimensional
regular ring and small-world networks, which allow us to explore the
crossover between the two limiting cases. We have obtained both
analytically and numerically the solution for relevant parameters of
the network and observed that our model exhibits the classical
phenomenon as that in the WS model. Our model may provide a useful
tool to investigate the influence of the clustering coefficient or
average path length in different dynamics processes taking place on
networks. In addition, using the idea presented here, one can also
construct models interpolating between homogeneous and heterogeneous
networks~\cite{GaMo06}.

\begin{acknowledgments}
This research was supported by the National Natural Science
Foundation of China under Grant Nos. 60373019, 60573183, and
90612007.
\end{acknowledgments}

\end{document}